\documentclass[12pt]{iopart}

\usepackage{graphicx}
\usepackage{dcolumn}
\usepackage{bm}

\usepackage{color}

\begin{document}

\title{Exact Results for the Bipartite Entanglement Entropy of the AKLT spin-1 chain}

\date{\today}

\author{Scott D. Geraedts}
\address{Department of Physics and Astronomy, McMaster University, Hamilton, ON, L8S 4L8, Canada}

\author{Erik S. S{\o}rensen}
\address{Department of Physics and Astronomy, McMaster University, Hamilton, ON, L8S 4L8, Canada}

\begin{abstract}
We study the bipartite entanglement between a sub-system of size $l$ and the rest of the system of total size $L$
as it occurs in a  spin-1 AKLT chain subject to open boundary conditions.
In this case the ground-state manifold is four-fold degenerate and there is strong dependence on the parity
of the number of spins, $L$.
We present exact analytical results for the von Neumann entanglement entropy,
as a function of both the size of the sub-system, $l$, and the total system size, $L$, for {\it all} four degenerate ground-states for both  
odd an even $L$.
In the large $l,L$ limit the entanglement entropy approaches $\ln(2)$  for the $S^z_T=\pm 1$ while
it approaches {\it twice} that value, $2\ln(2)$, for the $S^z_T=0$ states.
In all cases, it is found that this constant is approached exponentially fast defining a length scale
$\xi=1/\ln(3)$ equal to the known bulk correlation length.
\end{abstract}

\maketitle

\section{Introduction}
Quantum entanglement as it occurs in quantum spin chains is a property that has recently been under
intense study \cite{Sorensen,Hastings,Fan2,Calabrese,Michalakis}. The entanglement of a system can provide information
about the properties of that system, and long-distance entanglement is thought to be necessary for applications such as
quantum teleportation \cite{Boschi,Bouwmeester} and quantum cryptography. \cite{Gisin}
Exact results for the bipartite entanglement is from this perspective of considerable value.

One system displaying entanglement is that of an $S=1$ antiferromagnetic chain. \cite{Venuti} 
One generalized Hamiltonian for such a chain is given by:
\begin{equation}
H=J\sum_{i=1}^{L-1}[S_i\cdot S_{i+1}-\beta (S_i \cdot S_{i+1})^2],
\label{ham}
\end{equation}
where $\beta$ is a dimensionless parameter describing the biquadratic coupling. When $\beta=-1/3$, 
the system is at the Affleck-Kennedy-Lieb-Tasaki (AKLT) point, where the ground state of the system corresponds to a system where each $S=1$ spin is represented
as two $S=1/2$ spins, and $S=1/2$ spins from neighboring sites are combined into a singlet. \cite{AKLT1,AKLT} 
One and two-site entanglement at the AKLT point and for more generalized models
has been studied extensively with periodic boundary conditions.~\cite{Tribedi,Tribedi2,Hirano} 
When subject to periodic boundary conditions the hamiltonian, Eq.~(\ref{ham}), has
a nondegenerate singlet ground state at the AKLT point. \cite{Kennedy}
Some measures of entanglement has also been studied for the case of open boundary conditions~\cite{Fath,Fan,Katsura,Verstraete,Katsura2,Refael,Verstraete2,Alipour}, 
in this case the ground state of the system is four-fold degenerate~\cite{AKLT1,AKLT}, consisting of a singlet state, $S=0$, 
as well as a triplet state with $S=1,S_T^z=0,\pm1$. 
An interesting quantity to study is the bipartite entanglement entropy $S(l,L)$, 
the von Neumann entanglement of a subsystem of the chain with the rest of the chain:
\begin{equation}
S(l,L)\equiv -\hbox{Tr}\rho\log\rho,
\end{equation}
where $\rho$ is the reduced density matrix for the subsystem of size $l$ within the total system of length $L$.
Similar calculations have also been performed
for $S=\frac{1}{2}$ systems. \cite{Vidal,Laflorencie,Zhou} 
In physical systems well characterized as $S=1$ spin chains such as NENP~\cite{Renard} and Y$_2$BaNiO$_5$~\cite{DiTusa} the biquadratic term
is negligible, $\beta=0$, and impurities
likely cut the chains thereby effectively imposing open boundary conditions and restricting the length of such finite chain segments. 
The presence of the open boundaries has the peculiar effect of inducing $S=1/2$ excitations localized at the ends of the chain segment~\cite{Hagiwara}.
The physically most relevant point, $\beta=0$, is in the same phase as the AKLT point, the so called Haldane phase.
Within the Haldane phase, for $\beta\neq-1/3$, the four-fold ground-state degeneracy is lifted for finite $L$ and is replaced by an exponentially
low-lying triplet state above the singlet ground-state when the length of the system is even. For odd length systems the picture
is reversed and the triplet state is lowest. A complete characterization of the entanglement as it occurs for all 4 states in the
ground-state manifold would therefore be of interest. 
While $S(l,L)$ has been studied at the AKLT point for {\it periodic} boundary conditions by Hirano et al. \cite{Hirano}
the only result, by Alipour et al.~\cite{Alipour}, 
for the physically more interesting case of open boundary conditions is for the special case $l=1$
with $S_T^z=\pm 1$.
In the following, we present analytical results for the bipartite entanglement entropy for the AKLT system 
with {\it open} boundary conditions, as a function of both the size of the total system, $L$, and the size of the subsystem, $l$.
We explicitly present results for {\it all} four states for both even and odd length systems.

\section{Calculations}
In order to facilitate the calculations it is convenient to write the
ground-state wave function of Eq.~(\ref{ham}) in the following manner: \cite{AKLT1,AKLT,Kolezhuk,Arovas,Klumper}
\begin{equation}
|\Psi\rangle = \prod_{i} g_i, g_i=\left(\begin{array}{cc}
\frac{1}{\sqrt{2}}|0\rangle_i & -|+\rangle_i \\
|-\rangle_i & -\frac{1}{\sqrt{2}}|0\rangle_i \\
\end{array} \right).
\label{psi}
\end{equation}
In the above equation, $|0\rangle_i, |+\rangle_i$ and $|-\rangle_i$ 
are the states of the S=1 spin at site i. The matrix gives the four ground states of the system. 
The upper right and lower left entries correspond to the states with magnetization $S=1,S^z_T=1$ and $S=1,S^z_T=-1$. 
The two $S^z_T=0$ are as written not part of total spin multiplets and in light of the splitting of the ground-state
manifold away from the AKLT point it is therefore of interest to form total spin eigenstates which is conveniently done by defining:
\begin{equation}
\begin{array}{c}
|\Psi^0>_{singlet}=\frac{1}{\sqrt{2}}\left(\psi+SI(\psi)\right)\\
|\Psi^0>_{triplet}=\frac{1}{\sqrt{2}}\left(\psi-SI(\psi)\right)\\
\end{array}
\label{S=0}
\end{equation}
where $\psi$ is a diagonal entry in the wave function matrix, and $SI$ is the spin inversion operator.
We can use this wave function to calculate $S(l,L)$. 
First we break the spin chain up into two subchains, A and B. We denote the number of spins in A by $l$, compared to the total number of spins $L$. 
To find $S(l,L)$, we must first find the reduced density matrix, given by
\begin{equation}
\rho_{ij}=\sum_{j}a_{ij}^{\dag}a_{i'j},
\label{rho}
\end{equation}
where $i$ and $i'$ run over all possible configurations of the subsystem A, and $j$ runs over all possible configurations of the subsystem B. We can then 
diagonalize $\rho$ and compute $S(l,L)$. Given the simple matrix product form of the ground-states, Eq.~(\ref{psi}), it is possible to obtain explicit
expressions for $\rho$ using transfer matrix techniques. Though the reduced density matrix is large, 
it can be reduced to either a $2\times 2$ or $4\times 4$ matrix, reflecting the fact that the allowed states for the
subsystem A often is severely limited.

\begin{figure}[t]
\includegraphics[width=0.95 \linewidth,angle=0,clip]{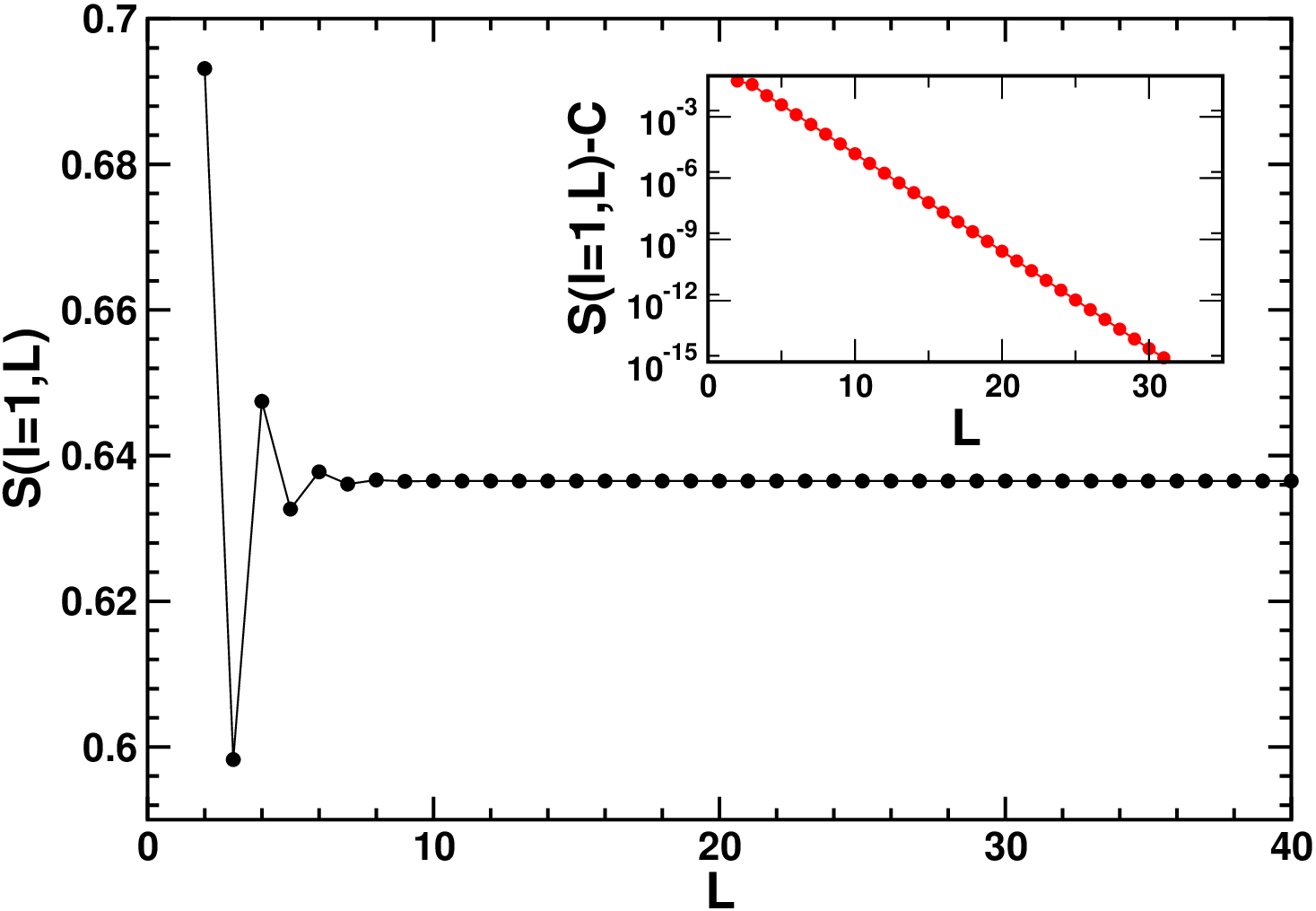}
\caption{The entanglement entropy of the system when $l=1$, as a function of $L$, for the $S=1,S_T^z=1$ state. The inset shows the convergence of the entropy towards its final value of $C=-(2/3)\ln(2/3)-(1/3)\ln(1/3)$. The entropy converges exponentially with a constant of $\ln(3)$, which implies a correlation length of $1/\ln(3)$ in this system.
}
\label{l1}
\end{figure}

\section{Results}
\subsection{$S(l=1,L)$, $L$ even and odd}
We begin our calculations by looking at the case where $l$, the size of subsystem A, is one. 
We have found that of the four degenerate ground states, 
two (the $S=1,S_T^z=\pm1$ states) have the same entanglement entropy by spin inversion symmetry.
We therefore have three different cases to consider.
\subsubsection{\it $S=1,S^z_T=\pm1$} 
In the following we explicitly consider the $S^z_T=1$ state of the triplet.
We use Eqn. ~(\ref{psi}) to find the wave function and subsequently the reduced
density matrix for each of the four ground states. 
We find that the reduced density matrix has only two eigenvalues, 
one of which is the probability, $x$, of a configuration containing the first spin (comprising all of subchain A) in a $|+\rangle$ state, 
and the remainder of the chain (subchain B) in a state with $S_{TB}^z=0$, 
and the other is the probability, $1-x$, of the first spin being in a $|0\rangle$ state and the remainder of the chain having $S_{TB}^z=1$. 
For the $S^z_T=1$ state configurations with the first spin in a $|-\rangle$ state does {\it not} occur at the AKLT point, hence, 
the probability, $x$ is simply the on-site magnetization of the first element of the chain, $\langle S^z_1\rangle$, which previously has
been determined for both {\it even} and {\it odd} $L$:\cite{Polizzi}
\begin{equation}
x=\frac{\frac{2}{3}-2(-3)^{-L}}{1-(-3)^{-L}}.\ \ \mathrm{(Any\ L)}
\label{x1}
\end{equation}
The reduced density matrix for the subsystem now takes the form:
\begin{equation}
\rho=\left[\begin{array}{ccc}
x & 0 \\
0 & 1-x \\
\end{array} \right].
\label{rho1}
\end{equation}
This leads to the final equation for the Von Neumann entropy:
\begin{equation}
S(l=1,L)=-x\ln(x)-(1-x)\ln(1-x),\ \ \mathrm{(Any\ L)}
\label{vn2}
\end{equation}
in agreement with the prior results by Alipour et al.\cite{Alipour} 
Interestingly, we have here related $S(l=1,L)$ directly to $\langle S^z_1\rangle$ an experimentally measurable quantity.
The result is a solution that converges exponentially fast with $L$ towards a final value of $C=-(2/3)\ln(2/3)-(1/3)\ln(1/3)$ as shown in Fig. \ref{l1}.
The exponential form allows for a determination of a length scale which from Eqs.~(\ref{x1}) and (\ref{vn2}) is seen to be $\xi=1/\ln(3)$ equal
to the known bulk correlation length of $1/\ln(3)$~\cite{Arovas} at the AKLT point.\\

\subsubsection{\it $S=0,S^z_T=0$} 
We now turn to a discussion of the first of the two $S^z_t=0$ states.
We note that the lower diagonal entry of Eq.~(\ref{psi}) 
is the spin inverse of the upper diagonal entry for even values of $L$. For odd $L$, the lower diagonal 
entry is the spin inverse of the upper diagonal entry times a factor of $-1$. This means that the singlet state when $L$ is even, 
and the $S^z_t=0$ triplet state when $L$ is odd, are given by the trace of Eq.~(\ref{psi}). 
In this case we find for $l=1$ simply a {\it constant} independent of L:
\begin{equation}
S(l=1,L)=\ln(3) \; \mbox{(L even)}.
\end{equation}
In this case the reduced density matrix is a 4x4 matrix with the following form:
\begin{equation}
\rho=\left[\begin{array}{cccc}
x & 0 & 0 & 0\\
0 & x & 0 & 0 \\
0 & 0 & x & 0\\
0 & 0 & 0 & 1-x \\
\end{array} \right].
\label{rho2}
\end{equation}
Where in the case of $l=1$, x is equal to $1/3$. 

In the case where $L$ is odd, the wave function is given by the upper left element of Eq.~(\ref{psi}) minus the lower right element. 
This yields the following expression for the entanglement entropy of the singlet state when $L$ is odd:
\begin{eqnarray}
x=&\frac{(1-(-3)^{-L+1})}{3(1-(-3)^{-L})}\ \ \mbox{(L odd)}\nonumber\\
S(l=1,L)&=-2x\ln(x)-(1-2x)\ln(1-2x).
\end{eqnarray}\\
In this case $S(l=1,L)$ now approaches the constant $\ln(3)$ exponentially with $L$.

\subsubsection{\it $S=1,S^z_T=0$} 
Using the same arguments as for the $S=0$ case, 
we see that for even $L$ the wave function is given by the upper left element of Eq.~(\ref{psi}) minus the lower right element, which produces the following result:
\begin{eqnarray}
x=&\frac{(1-(-3)^{-L+1})}{3(1-(-3)^{-L})}\ \ \mathrm{(L\ even)}\nonumber\\
S(l=1,L)&=-2x\ln(x)-(1-2x)\ln(1-2x),
\end{eqnarray}
the same result we found for {\it odd} $L$ for the $S=0$ state. The reduced density matrix is still a 4x4 matrix, but it has the form:
\begin{equation}
\rho=\left[\begin{array}{cccc}
x & 0 & 0 & 0\\
0 & x & 0 & 0 \\
0 & 0 & y & 0\\
0 & 0 & 0 & 1-2x-y \\
\end{array} \right].
\label{rho3}
\end{equation}
Where for $l=1$, $y=0$. 

In the present case, for odd $L$,  the wave function is given by the trace of Eq.~(\ref{psi}), so 
the entanglement entropy is given by the formula of Hirano et al.:
\begin{equation}
S(l=1,L)=\ln(3) \; \mbox{(L odd)}
\end{equation}

To summarize, we have found that $S(l=1,L)$ approaches either $-(2/3)\ln(2/3)-(1/3)\ln(1/3)$ or $\ln(3)$
in some cases in an exponential manner with $L$ in other cases the result $S(l=1,L)$ is independent of $L$.

\begin{figure}[t]
\includegraphics[width=0.95 \linewidth,angle=0,clip]{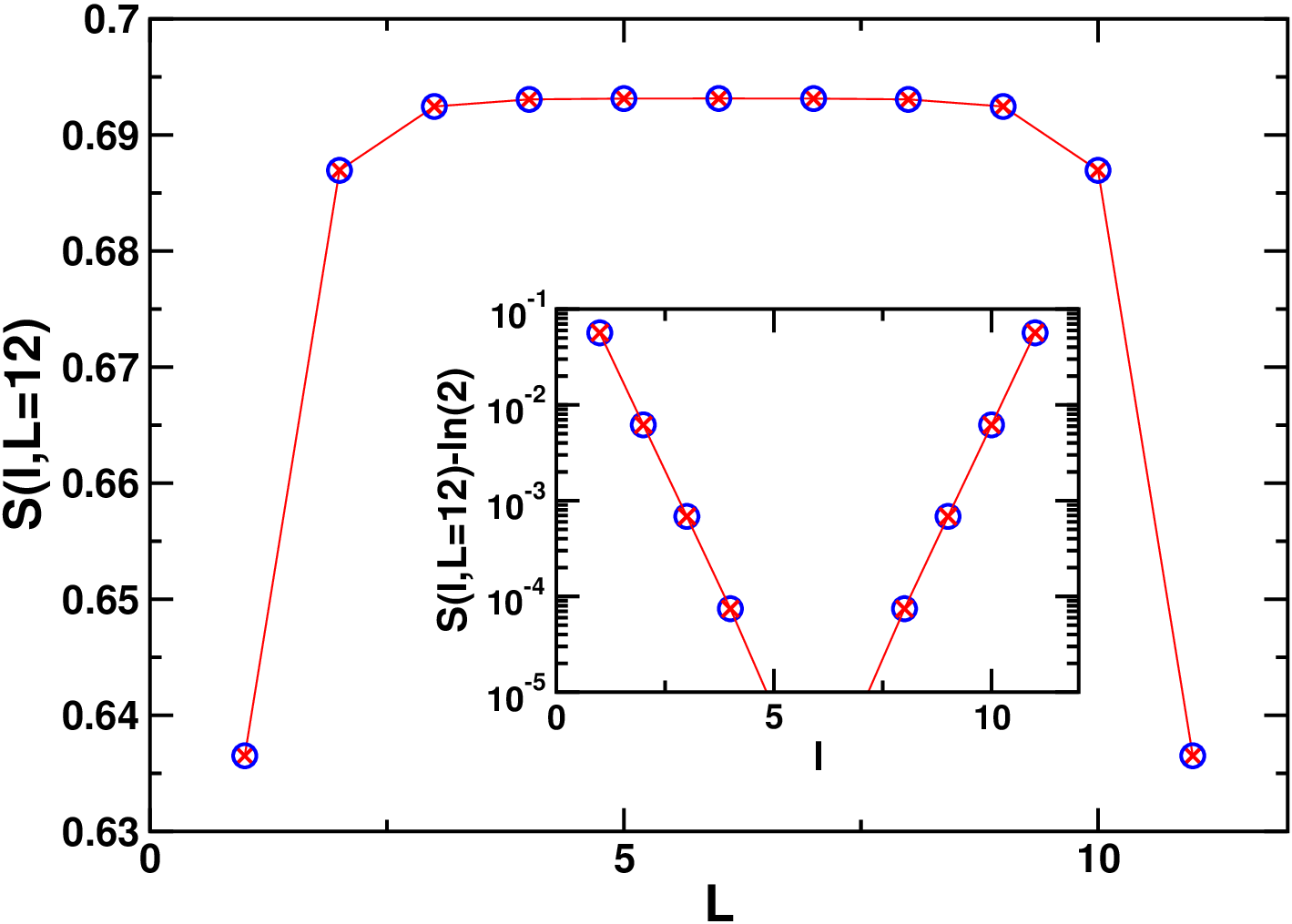}
\caption{The entanglement entropy for the $S=1,S^z_T=1$ state as a function of $l$ when $L=12$ (Eq. ~(\ref{x2})). 
The open circles indicate the corresponding exact diagonalization results used as a check. The
  inset shows the convergence of our data towards its final value of $\ln(2)$.
  The entropy therefore converges exponentially as $l\rightarrow L/2$.
  \label{l12} }
  \end{figure}

\subsection{$S(l, L)$, $L$ even}

We now generalize our results to any size $l$ of the subsystem A, with $L$ even.
\subsubsection{\it $S=1,S_T^z=\pm1$} 
Since the entanglement is the same for $S_T^z=\pm1$ we in the following take $S^z_T=1$.
In the case of $S_T^z=1$, we find that $\rho$ again has two eigenvalues. 
One is the probability of finding the system in a state such that the total
magnetization of subsystem A $S_{T,A}^z=1$, and the total magnetization of subsystem B $S_{T,B}^z=0$. 
The other is the opposite case: $S_{T,A}^z=0, S_{T,B}^z=1$. If we denote the first eigenvalue by x, 
then the Von Neumann entropy is given by Eq.~(\ref{vn2}), and the reduced density matrix has form of Eq.(\ref{rho1}) The value x at any $l$ is given by:
\begin{eqnarray}
x=&\frac{(1-(-3)^{-l})(1+(-3)^{-L+l})}{2(1-(-3)^{-L})}\ \mathrm{(L\ even)}\nonumber\\
S(l,L)&=-x\ln(x)-(1-x)\ln(1-x).
\label{x2}
\end{eqnarray}
This result is plotted as a function of $l$ when $L=12$ in Fig. \ref{l12}.
At $L=2l$, this gives a result of {\it exactly} $S(l=L/2,L)=\ln(2)$ independent of $L$. We also note that when $l$ and $L$ are both large, $S(l,L)$
converges to $\ln(2)$
again in an exponential manner on a length scale of $1/\ln(3)$. This asymptotic value of the entanglement entropy seems natural since
in the present case the boundary of the subsystem A will cut a single valence bond resulting in a contribution of $\ln(2)$ to the entanglement entropy.

\subsubsection{\it $S=0,S^z_T=0$} 
In the case where $L$ is even, the wave function is given by the trace 
of Eq.~(\ref{psi}), so the reduced density matrix is given by Eq.(\ref{rho2}) and the entanglement entropy is given by the following formula: \cite{Hirano}
\begin{eqnarray}
x=&\frac{(1-(-3)^{-l})(1-(-3)^{-L+l})}{4(1-(-3)^{-L+1})}\ \mathrm{(L\ even)}\nonumber\\
S(l,L)&=-3x\ln(x)-(1-3x)\ln(1-3x).
\label{x3}
\end{eqnarray}
Two of the degenerate eigenvalues correspond to the probabilities of subsystem A having $S_{T,A}^z=\pm1$, and the sum of the
remaining degenerate eigenvalue and the non-degenerate eigenvalue is the probability of subsystem A having $S_{T,A}^z=0$.
Strikingly, in this case
the entanglement entropy quickly approaches $2\ln(2)$ for large $l,L$, {\it twice} the result for the
$S=1,S_T^z=\pm1$ states. As above we can argue that this asymptotic value of the entanglement entropy is natural
since in addition to cutting a {\it single} valence bond at the boundary, the subsystem now also cuts the singlet
formed by the two effective $S=1/2$ chain boundary excitations resulting, in a contribution of $2\ln(2)$ to the 
entanglement entropy.
It is possible to argue that the result, Eq.~(\ref{x3}), is independent of the boundary conditions and the above result does
agree with previous results for {\it periodic} boundary conditions.~\cite{Hirano}

\begin{figure}[t]
\includegraphics[width=0.95 \linewidth,angle=0,clip]{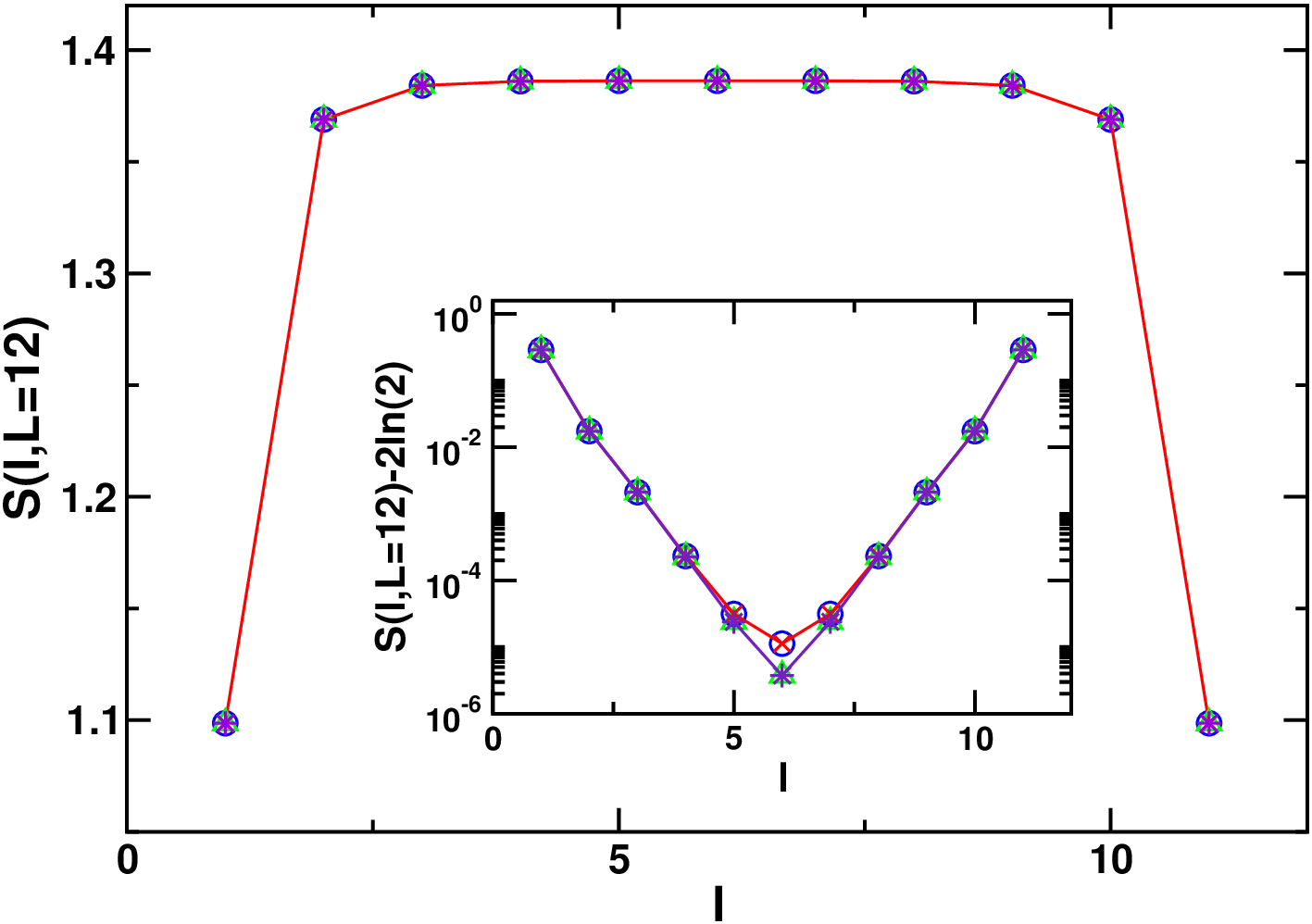}
\caption{The entanglement entropy for the $S=0,S^z_T=0$ state (red) and the
  $S=1,S^z_T=0$ state (purple) as a function of $l$ when $L=12$ (Eq.
      ~(\ref{x2})). (color online) The open symbols indicate the corresponding
    exact diagonalization results used as a check. The inset shows the convergence of the
    analytical data to the final value of $2\ln(2)$. 
    \label{l12_2} }
    \end{figure}
\subsubsection{\it $S=1,S^z_T=0$} 
In this case 
the reduced density matrix now has four eigenvalues and is given by Eq.(\ref{rho3}). 
The two degenerate eigenvalues correspond to the probability of subsystem A being in
a state with $S_{T,A}^z=\pm1$, 
The sum of the non-degenerate eigenvalues is the probability that $S_{T,A}^z=0$. 
We then find the following equation:
\begin{eqnarray}
x=&\frac{(1-(-3)^{-l})(1-(-3)^{-L+l})}{4(1-(-3)^{-L})}\ \mathrm{(L\ even)}\nonumber\\
y=&\frac{(1-(-3)^{-l})(1-(-3)^{-L+l+1})}{4(1-(-3)^{-L})}\nonumber\\
S(l,L)&=-2x\ln(x)-y\ln(y)\nonumber\\
 &-(1-2x-y)\ln(1-2x-y).
\label{x4}
\end{eqnarray}
Again we observe that the asymptotic value of the entanglement entropy is $2\ln(2)$. Hence, we
have explicitly showed that the three states of the triplet $S=1,S^z_T=0,\pm1$ only the two $S^z_T=\pm1$
states related by spin inversion yield  the same entanglement entropy while the $S=1,S^z_T=0$ state not
only differs by an overall factor of $\ln(2)$ but also in subleading terms.
The results Eqs.~(\ref{x3}) and (\ref{x4}) are plotted in Fig. \ref{l12_2} where they are compared with
exact diagonalization results. 
Recent work \cite{Pollmann} has suggested that the Haldane phase is characterized by two-fold degeneracy in the eigenvalues of the reduced density matrix of the ground state, which is consistent with these findings.

\subsection{$S(l,L)$, $L$ odd}
As before when we detailed the $S(l=1,L)$ case we expect rather strong dependence on the parity of $L$
for the general $S(l,L)$.
For completeness we now give the equations for the entanglement entropy also in this case.
\subsubsection{\it $S=1,S^z_T=\pm1$} 
In this case there is no dependence on the parity of $L$ and the result is the
same as for even $L$ given in Eq.~(\ref{x2}).

\subsubsection{\it $S=0,S^z_T=0$} 
By the same arguments as the $l=1$ case, the entanglement entropy for odd $L$ is given by:
\begin{eqnarray}
x=&\frac{(1-(-3)^{-l})(1-(-3)^{-L+l})}{4(1-(-3)^{-L})}\ \mathrm{(L\ odd)}\nonumber\\
y=&\frac{(1-(-3)^{-l})(1-(-3)^{-L+l+1})}{4(1-(-3)^{-L})}\nonumber\\
S(l,L)&=-2x\ln(x)-y\ln(y)\nonumber\\
&-(1-2x-y)\ln(1-2x-y).
\label{x5}
\end{eqnarray}

\subsubsection{\it $S=1,S^z_T=0$} 
Similarly, the wave function for $S=1,S^z_T=0$, L odd is given by the trace of Eq.~(\ref{psi}), and so the entanglement entropy is:
\begin{eqnarray}
x=&\frac{(1-(-3)^{-l})(1-(-3)^{-L+l})}{4(1-(-3)^{-L+1})}\ \mathrm{(L\ odd)}\nonumber\\
S(l,L)&=-3x\ln(x)-(1-3x)\ln(1-3x).
\label{x6}
\end{eqnarray}

\section{Conclusions}
We have obtained explicit analytical equations for the bipartite entanglement entropy of a spin-1
chain at the AKLT point for all four states of the ground-state manifold. For the case where 
$S=1,S^z_T=\pm 1$, we have found that for large system
sizes the entanglement entropy approaches $\ln(2)$ while for the $S=0,S^z_T=0$ and $S=1,S^z_T=0$ cases, the
entanglement entropy approaches $2\ln(2)$. Hence, the entanglement entropy is in this case not $SU(2)$ invariant. 
In all cases where there is an explicit $l$ or $L$ dependence have we found that the asymptotic value is approached
in an exponential manner defining a length scale of $\xi=1/\ln(3)$ equal to the bulk correlation length. 
Also of interest are the time-reversal and spin-reversal invariant states $|\phi \rangle=|+1\rangle \pm |-1\rangle$. Exact calculations of the entanglement entropy 
of these states for small $L$ suggests that their entanglement entropy converges towards $2\ln(2)$, which is consistent with 
the interpretation that each factor of $\ln(2)$ corresponds to a cutting of a bond. We have so far been unable to obtain an explicit 
formula for the entanglement entropy of these states. \\

\vspace{0.5in}
\bibliographystyle{ieeetr}
\bibliography{refs}

\end{document}